\documentclass[preprint,tightenlines,showpacs,showkeys,floatfix,
nofootinbib,superscriptaddress,fleqn]{revtex4}
\usepackage{bm}
\usepackage{dcolumn}
\usepackage{amsmath}
\usepackage{amsfonts}
\usepackage{amssymb}
\usepackage{amscd}
\usepackage{amsthm}
\usepackage{graphicx}

\begin{document}

\preprint{PNU-NTG-09/2007}
\title{$1/N_c$ corrections to the magnetic susceptibility \\
of the QCD vacuum}
\author{Klaus Goeke}
\email{Klaus.Goeke@tp2.rub.de}
\affiliation{Institut f\"ur Theoretische Physik II,
Ruhr-Universit\" at Bochum, D--44780
Bochum, Germany}
\author{Hyun-Chul Kim}
\email{hchkim@pusan.ac.kr}
\affiliation{Department of Physics and Nuclear Physics \& Radiation
  Technology Institute 
(NuRI), Pusan National University, 609-735 Busan, Republic of Korea}
\author{M.M. Musakhanov}
\email{musakhanov@pusan.ac.kr}
\affiliation{Theoretical Physics Department, National University of
  Uzbekistan, Tashkent 
700174, Uzbekistan}
\author{Marat Siddikov}
\email{Marat.Siddikov@tp2.rub.de}
\affiliation{Institut f\"ur Theoretische Physik II, Ruhr-Universit\"
  at Bochum, D--44780 
Bochum, Germany}
\affiliation{Theoretical Physics Department, National University of
  Uzbekistan, Tashkent 
700174, Uzbekistan}
\date{August 2007}

\begin{abstract}
We investigate the magnetic susceptibility of the QCD vacuum with
the $1/N_c$ corrections taken into account, based on the instanton
vacuum.  Starting from the instanton liquid model we derive the
gauged light-quark partition function in the presence of the
current quark mass as well as of external Abelian vector and
tensor fields.  We consider the $1/N_c$ meson-loop corrections
which are shown to contribute to the magnetic susceptibility by
around $15\,\%$ for the up (and down) quarks. We also take into
account the tensor terms of the quark-quark interaction from the
instanton vacuum as well as the finite-width effects, both of
which are of order $\mathcal{O}(1/N_c)$.  The effects of the
tensor terms and finite width turn out to be negligibly small. The
final results for the up-quarks are given as: 
$\chi\langle i\psi^\dagger \psi\rangle_0 \simeq 35-40
\,\mathrm{MeV}$ with the quark condensate $\langle i\psi^\dagger
\psi\rangle_0$.  We also discuss the pion mass dependence of the
magnetic susceptibility in order to give a qualitative guideline
for the chiral extrapolation of lattice data.
\end{abstract}


\maketitle


\section{Introduction}

The QCD vacuum is known to be one of the most complicated states
entangled with perturbative and strong nonperturbative
fluctuations. In particular, the condensates of the quarks and
gluons reveal their non-perturbative aspect and character. The
quark condensate is taken as an order parameter associated with
spontaneous chiral symmetry breaking (S$\chi $SB) which is
probably one of the most important quantities for low-energy
hadronic phenomena. Actually, the presence of an external source
allows a more profound study of the QCD vacuum. For example, the
constant electromagnetic field $F_{\mu \nu }$ induces another type
of condensates leading to the nonzero magnetic susceptibility
$\chi$ of the QCD vacuum defined as:
\begin{equation}
\langle \psi _{f}^{\dagger }\sigma _{\mu \nu }\psi _{f} \rangle
_{F}=e_{f}\,\chi _{f}\,\langle i\psi _{f}^{\dagger }\psi
_{f}\rangle _{0}\,F_{\mu \nu },  \label{eq:chi}
\end{equation}
where $e_{f}$ denotes the quark electric charge with the
corresponding flavor $f$. It is natural to have the quark
condensate $\langle i\psi _{f}^{\dagger }\psi _{f}\rangle _{0}$ in
the chiral limit as a normalization factor in the right-hand side
of Eq.(\ref{eq:chi}), since S$\chi $SB is  responsible for these
quantities, i.e. the quark condensate and magnetic susceptibility.
The value of $\chi _{f}$ was already predicted in the QCD sum rule
and vector-dominance model~\cite{Ioffe:1983ju,Belyaev:ic,Balitsky:aq,
Ball:2002ps}: $\chi _{u}\langle i\psi _{u}^{\dagger }\psi
_{u}\rangle _{0}=40\sim 70\,\mathrm{MeV}$ at the scale of
$1\,\mathrm{GeV}$, and in the instanton liquid model for the QCD
vacuum~\cite{Petrov:1998kg} to $\sim 38\,\mathrm{MeV}$ and in Ref.
~\cite{Kim:2004hd} to $45\sim 50\,\mathrm{MeV}$.\footnote{In the
following we give all numbers for the up-quarks, which are due to
assumed isospin invariance identical to the numbers for the down
quarks. Thus we ignore the index $u$ or $f$.} Moreover,
Ref.~\cite{Braun:2002en} suggested that the magnetic
susceptibility $\chi$ may be measured in the exclusive
photoproduction of hard dijets $\gamma +N\rightarrow
(\bar{q}q)+N$.

In the present work, we want to extend the previous
work~\cite{Kim:2004hd}, incorporating the $1/N_c$
corrections~\cite{Kim:2005jc,Goeke:2007bj}. Since the instanton
vacuum explains S$\chi$SB naturally via quark zero modes, it may
provide a good framework to study the $\chi$ of the light-quark
vacuum. Moreover, there are only two parameters in this approach,
namely the average instanton size $\bar \rho\approx \frac{1}{3}\,
\mathrm{fm}$ and average inter-instanton distance $\bar R\approx
1\, \mathrm{fm}$. The normalization scale of this approach can be
defined by the average size of instantons and is approximately
equal to $\rho^{-1}\approx 0.6\, \mathrm{GeV}$. The values of
the $\bar \rho$ and $\bar R$ were estimated many years ago
phenomenologically in Ref.~\cite{Shuryak:1981ff} as well as
theoretically in
Ref.~\cite{Diakonov:1983hh,Diakonov:2002fq,Schafer:1996wv}. 
Furthermore, it was confirmed by various lattice simulations of
the QCD vacuum \cite{Chu:vi,Negele:1998ev,DeGrand:2001tm}. Also
lattice calculations of the quark propagator~\cite%
{Faccioli:2003qz,Bowman:2004xi} are in a remarkable agreement with
that of Ref.~\cite{Diakonov:1983hh}. A recent lattice simulation
with the interacting instanton liquid model obtains $\bar
\rho\simeq 0.32\,\mathrm{fm}$ and $\bar R\simeq 0.76
\,\mathrm{fm}$ with the finite current quark mass $m$ taken into
account~\cite{Cristoforetti:2006ar}. Considering the $1/N_c$
corrections, Ref.~\cite{Goeke:2007bj} found that the values of the
instanton size and average distance should be readjusted as
\begin{equation}
\bar \rho=0.35\,\mathrm{fm},\;\;\;\bar R=0.86\,\mathrm{fm}
\label{eq:rhoR}
\end{equation}
with $F_\pi =88\,\mathrm{MeV}$ and
$\langle i\psi^{\dagger }\psi\rangle _{0} =(255\,\mathrm{MeV%
})^3$ required in the chiral limit. The values in
Eq.(\ref{eq:rhoR}) will be used in the following calculations.

In order to consider the $1/N_c$ corrections to the magnetic
susceptibility of the QCD vacuum, we closely follow the formalism
developed in Refs.~\cite{Kim:2005jc,Goeke:2007bj}. In order to
calculate the magnetic susceptibility, we first have to find the
light-quark partition function
$%
Z[V,T,m]$ in the presence of the Abelian external vector $V_\mu$
and antisymmetric tensor $T_{\mu\nu}$ fields. So, we first
calculate a zero-mode approximated quark propagator $\tilde S$ in
the instanton ensemble and in the presence of these external
fields. Using this propagator, we are able to
compute the low-frequency part of the quark determinant for the
non-zero $m$ with the external Abelian vector $V$ and tensor $T$
fields. The relevant techniques have been already developed in
Refs.~\cite{Kim:2004hd}. 

The smallness of the packing parameter $\pi(\frac{\bar \rho}{\bar
R})^4\approx 0.1$ makes it possible to average the determinant
over collective coordinates of instantons with fermionic
quasiparticles, i.e. constituent quarks $\psi$ introduced. The
averaged determinant turns out to be the light-quark partition
function $Z[V,T,m]$ which is a functional of $V$ and $T$ and can
be represented by a functional integral over the constituent quark
fields with the gauged effective chiral action
$S[\psi^\dagger,\psi,V,T]$.  However, it is not trivial to make
the action gauge-invariant due to the nonlocality of the
quark-quark interactions generated by instantons. In the previous
paper~\cite{Musakhanov:2002xa}, it was demonstrated how to gauge
the nonlocal effective chiral action in the presence of the
external electromagnetic field and was shown that the low-energy
theorem of the axial anomaly relevant to the process $G\tilde
G\rightarrow \gamma\gamma$ is satisfied (see also Refs.
\cite{Musakhanov:1996qf,Salvo:1997nf}).  The gauged effective chiral
action was also successfully applied to various observables of mesons
and vacuum properties~\cite{Kim:2004hd,Nam:2006sx,Ryu:2006bf,
Nam:2007fx}.  

The present work is organized as follows: In Section II, we review
the general formalism for the magnetic susceptibility of the QCD
vacuum. We first show how to derive from the instanton vacuum the
gauged effective chiral action in the presence of the current
quark mass as well as of external vector and tensor fields. We
then derive the meson propagator for the meson-loop corrections to
the magnetic susceptibility of the QCD vacuum, considering the
fluctuation around the saddle-point.  In Section III, we discuss
the meson-loop corrections to the magnetic susceptibility, the
contribution of the tensor quark-quark interactions to it, and the
effects of the finite width of the instanton size, all of which
are $1/N_c$ corrections to the magnetic susceptibility.  In
Section IV, we present the final results of the magnetic
susceptibility of the QCD vacuum and discuss its relevance to the
chiral extrapolation in lattice QCD, extending the results with
larger pion masses employed. In the final Section, we summarize
the present work and draw conclusions.

\section{General Formalism}
\subsection{Light quark partition function in the presence of
external vector and tensor fields}
The light quark  partition function $Z[V,T,m]$ with the quark mass
$m$ and  external vector $V_\mu $ and tensor $T_{\mu\nu}$ fields is defined as
\begin{eqnarray}
\label{GenFunc} Z[V,T,m]=\int D\,A_\mu\,e^{-\frac{1}{4}G^2}
\mathrm{Det}(i\rlap{/}{\partial} +\rlap{/}{A} + im +\rlap{/}{V} +
\sigma_{\mu\nu} T_{\mu\nu}),
\end{eqnarray}
where $A_\mu$ is the gluon field and $G_{\mu\nu}$ is the gluon field
strength tensor.  The basic assumption of the instanton liquid model
is that one can evaluate the integral in a quasi-classical
approximation, expanding it around the classical vacuum.  The first
evaluation of the partition function Eq.(\ref{GenFunc}) was performed
in Ref.~\cite{Diakonov:1985eg,Diakonov:1995qy} in the absence 
of the external fields and in the chiral limit.  The main purpose of
the present paper is the extension of those results to the case of
the nonzero quark mass $m$ and external $V,T$ fields, following the
method given in \cite{Kim:2004hd,Goeke:2007bj}.  For this we split the
quark determinant into the low- and high-frequency parts according 
${\mathrm{Det}}={\mathrm{Det}}_{\mathrm{low}}{\mathrm{Det}}_{\mathrm{high}}$
and concentrate on the evaluation of ${\mathrm{Det}}_{\mathrm{low}}$,
which is responsible for the low-energy domain.  The high-energy part
${\mathrm{Det}}_{\mathrm{high}}$ is responsible
mainly for the perturbative coupling renormalization.

We start with the zero-mode approximation for the propagator of a
quark interacting with the $i$-th instanton. This propagator was
considered
in~\cite{'tHooft:1976fv,Lee:sm,Diakonov:1985eg,Diakonov:1995qy}:
\begin{equation}  \label{SiDP}
S_i=\frac{1}{\rlap{/}{p} + \rlap{/}{A}_i
  +im}=\frac{1}{\rlap{/}{p}} +
\frac{|\Phi_{i,0}\rangle\langle\Phi_{i,0}|}{im}.
\end{equation}
While this zero approximation is good for small values of the
current quark mass $m$, we need to extend it beyond the chiral
limit as proposed in our previous
works~\cite{Musakhanov:1998wp,Musakhanov:vu, Musakhanov:2002xa,
Kim:2004hd} as follows:
\begin{equation}  \label{Si}
S_i=S_{0} + S_{0}\rlap{/}{p}\frac{|\Phi_{0i}\rangle\langle \Phi_{0i}|}{c_i} %
\rlap{/}{p} S_{0} ,
\end{equation}
where
\begin{equation}
c_i=-\langle\Phi_{0i}|\rlap{/}{p} S_{0} \rlap{/}{p} |\Phi_{0i}\rangle = i%
m\langle\Phi_{0i}|S_{0}\rlap{/}{p} |\Phi_{0i}\rangle = im%
\langle\Phi_{0i}|\rlap{/}{p} S_{0}|\Phi_{0i}\rangle.
\end{equation}
The approximation given in Eq.(\ref{Si}) allows us to project
$S_i$ to the correct zero-modes with the finite $m$:
\begin{equation}
S_i|\Phi_{0i}\rangle = \frac{1}{im}|\Phi_{0i}\rangle,\,\,\,
\langle\Phi_{0i}|S_i =\langle\Phi_{0i}|\frac{1}{im}.
\end{equation}

We can write the total quark propagator $\tilde{S}$ in the
presence of the whole instanton ensemble $A$ and external vector
($V$) and tensor ($T$) fields, and the quark propagator with a
single instanton $A_i$ as well as $V$ and $T$ as follows:
\begin{equation}
\tilde S = \frac{1}{\rlap{/}{p} + \rlap{/}{A} +\rlap{/}{V} + \sigma_{\mu\nu} T_{\mu\nu} +i%
m},\;\; \tilde S_i= \frac{1}{\rlap{/}{p} + \rlap{/}{A_i}
+\rlap{/}{V} + \sigma_{\mu\nu} T_{\mu\nu} +im}.
\end{equation}
Here, we assume that the total instanton field $A$ may be
approximated as a sum of the single instanton fields,
$A=\sum_{i=1}^N A_i$, which is justified with the values of $\bar
\rho$ and $\bar R$ in Eq.(\ref{eq:rhoR}). With the instanton
fields turned off, we write the quark propagator in the presence
of the external fields and free quark propagator as follows:
\begin{equation}
\tilde S_0=\frac{1}{\rlap{/}{p} + \rlap{/}{V} +\sigma_{\mu\nu}
T_{\mu\nu}+im}, \;\;\; S_0=\frac{1}{\rlap{/}{p} +im}.
\end{equation}
We now expand the total quark propagator $\tilde S$ with respect
to the single instanton field $A_i$:
\begin{equation}
\tilde S=\tilde S_0+\sum_i (\tilde S_i-\tilde S_0)+\sum_{i\not=j} (\tilde
S_i-\tilde S_0)\tilde S^{-1}_0(\tilde S_j-\tilde S_0)+\cdots .  \label{S-tot}
\end{equation}
The next step is to expand $\tilde{S}$ with respect to the external fields $V
$ and $T$,
and express $\tilde S_i$ in terms of $S_i$. Since we use the zero-mode approximation,
the expansion with respect to the vector field breaks the gauge invariance.  In
order to restore it, we introduce the following auxiliary field $V^{\prime }$ and
gauge connection $L_i$:
\begin{equation}
\rlap{/}{V}_i^\prime=\bar L_i(\rlap{/}{p}+\rlap{/}{V})L_i-\rlap{/}{p}.
\end{equation}
The gauge connection $L_i$ is defined as a path-ordered exponent
\begin{eqnarray}
L_i(x,z_i)&=&\mathrm{P} \exp\left(i\int_{z_i}^x dy_\mu V_\mu(y)\right),
\notag \\
\bar L_i(x,z_i)&=&\gamma_0 L_i^\dagger(x,z_i)\gamma_0,  \label{transporter}
\end{eqnarray}
where $z_i$ denotes an instanton position. The field $V^{\prime }_i(x,z_i)$
under flavor rotation $\psi(x)\rightarrow U(x)\psi(x)$ is transformed as $%
V^{\prime }_i(x,z_i)\rightarrow U(z_i)V^{\prime }_i(x,z_i)U^{-1}(z_i)$. The
propagators $\tilde S_i$ and $\tilde S_0$ then have the following form:
\begin{eqnarray}
\tilde S_i &=& L_iS^{\prime }_{i}\bar L_i,\;\; S^{\prime }_{i}=\frac{1}{%
\rlap{/}{p} +\rlap{/}{A_i} +\rlap{/}{V_i^{\prime
}}+\sigma_{\mu\nu} T_{\mu\nu}+im},
\cr \tilde S_0 &=& L_iS^{\prime }_{0i}\bar L_i,\;\; S^{\prime }_{0i}=\frac{1%
}{\rlap{/}{p} + \rlap{/}{V_i^{\prime }}+\sigma_{\mu\nu}
T_{\mu\nu}+im},
\end{eqnarray}
Expanding $S^{\prime }_{i}$ with respect to $\rlap{/}{V}_{i}^{\prime }$ and
resumming it, we get
\begin{equation}
S^{\prime }_{i} = S_i(1+\sum_n (-\hat V_{i}^{\prime }S_i)^n) =S^{\prime
}_{0i} + S^{\prime }_{0i}\rlap{/}{p}\frac{|\Phi_{0i}\rangle\langle \Phi_{0i}|%
}{c_i - b_i} \rlap{/}{p} S^{\prime }_{0i} ,  \label{Si'}
\end{equation}
where
\begin{eqnarray}  \label{bi}
b_i &=& \langle\Phi_{0i}|\rlap{/}{p} (S^{\prime }_{0i}-S_0) \rlap{/}{p}
|\Phi_{0i}\rangle, \\
c_i - b_i &=&-\langle\Phi_{0i}| \rlap{/}{p} S^{\prime }_{0i} \rlap{/}{p}%
|\Phi_{0i}\rangle.  \notag
\end{eqnarray}
Rearranging Eq.(\ref{S-tot}) for the total propagator, we obtain
\begin{eqnarray}
\tilde S&=&\tilde S_{0} + \tilde S_{0}\sum_{i,j} \bar L_i^{-1} \rlap{/}{p}
|\Phi_{i0}\rangle\left(\frac{1}{-\mathcal{D}}+\frac{1}{-\mathcal{D}}
\mathcal{C} \frac{1}{-\mathcal{D}} + \ldots\right)_{ij} \langle\Phi_{0j}|%
\rlap{/}{p} L_j^{-1}\tilde S_{0}\cr &=&\tilde S_{0} + \tilde
S_{0}\sum_{i,j}\bar L_i^{-1}\rlap{/}{p} |\Phi_{i0}\rangle \left(\frac{1}{-%
\mathcal{V}-\mathcal{T}}\right)_{ij} \langle\Phi_{0j}|\rlap{/}{p}
L_j^{-1}\tilde S_{0} ,  \label{propagator}
\end{eqnarray}
where
\begin{eqnarray}
\mathcal{V}_{ij}&=&\langle\Phi_{0i}|\rlap{/}{p} (L_i^{-1}\tilde S_{0}\bar
L^{-1}_j) \rlap{/}{p}|\Phi_{0j}\rangle -\langle\Phi_{0i}|\rlap{/}{p} S_{0}
L_i^{-1}L_j \rlap{/}{p} |\Phi_{0j}\rangle,  \notag \\
\mathcal{T}_{ij}&=&(1-\delta_{ij})\langle\Phi_{0i}|\rlap{/}{p} S_0
L_i^{-1}L_j\rlap{/}{p} |\Phi_{0j}\rangle,  \notag \\
\mathcal{D}_{ij}&=&\delta_{ij} \mathcal{V}_{ij} \equiv (b_i-c_i) \delta_{ij},%
\cr \mathcal{C}_{ij}&=&(1-\delta_{ij})\mathcal{V}_{ij}.
\end{eqnarray}
We introduce now the modified zero-mode solution:
\begin{equation}
|\phi_0\rangle=\frac{1}{\rlap{/}{p}}L \rlap{/}{p} |\Phi_0\rangle,
\end{equation}
which has the same chiral properties as the zero-mode solution $%
|\Phi_0\rangle$.
Then we get
\begin{equation}
\tilde S -\tilde S_{0} = -\tilde S_{0}\sum_{i,j}\rlap{/}{p} | \phi_{0i}
\rangle \langle\phi_{0i}|\left(\frac{1}{\mathcal{V}+\mathcal{T}}\right)|
\phi_{0j} \rangle \langle\phi_{0j}|\rlap{/}{p} \tilde S_{0}
\label{propagator1}
\end{equation}
with
\begin{equation}
\mathcal{V}+\mathcal{T}=\rlap{/}{p} \tilde S_{0}\rlap{/}{p}.
\end{equation}
The final explicit form for Eq.(\ref{propagator1}) is written as
\begin{equation}
\mathrm{Tr} (\tilde S -\tilde S_0 )= -\sum_{i,j}\langle\phi_{0,j}|\rlap{/}{p}%
\, ({\tilde S_{0}}^2)\, \rlap{/}{p} |\phi_{0,i}\rangle\langle\phi_{0,i}|%
\left(\frac{1}{ \rlap{/}{p}\tilde S_{0}\rlap{/}{p}}\right)|\phi_{0,j}\rangle.
\end{equation}

In order to derive the low-frequency part of the quark
determinant, we now introduce a matrix operator $\tilde{B}(m)$ defined
as follows:
\begin{equation}
\tilde B(m)_{ij} = \langle\phi_{0,i}|(\rlap{/}{p} \tilde S_0 \rlap{/}{p}
)|\phi_{0,j}\rangle = \langle\Phi_{0i}|\rlap{/}{p}
\left(L_{i}^{-1} \,\tilde S_{0}\bar L_{j}^{-1}\right)\rlap{/}{p}
|\Phi_{0j}\rangle,
\end{equation}
where $i,j$ are indices for the different instantons. Then, we can
show that
\begin{eqnarray}
\ln{\left(\mathrm{Det}_{\mathrm{low}}\right)}& = & \mathrm{Tr} \ln{\left(
\frac{i\rlap{/}{\partial} + \rlap{/}{A} + \rlap{/}{V}+ \sigma_{\mu\nu}
  T_{\mu\nu} +i\hat{m}}{ i\rlap{/}{\partial}+\rlap{/}{V}+
  \sigma_{\mu\nu} T_{\mu\nu}+ im} \right)} \\
&=& i \mathrm{Tr} \int^m dm^{\prime }(\tilde S(m^{\prime })- \tilde
S_{0}(m^{\prime }))\cr &=& \sum_{i,j} \int^m dm' \frac{d \tilde B(m^{\prime
})_{ij}}{dm'}(\tilde B(m'))_{ji}^{-1} =\mathrm{Tr} \ln {\tilde
B(m)},  \notag
\end{eqnarray}
 where $\tilde{\mathrm{Tr}}$ denotes the trace over the subspace of the quark zero modes.
Thus, we have
\begin{equation}
{\mathrm{Det}}_{\mathrm{low}} [V,T,m] \cong \mathrm{Det} \tilde
B(m),  \label{tildeB}
\end{equation}
where $\tilde B$ is the extension of Lee-Bardeen matrix
$B$~\cite{Lee:sm} in the presence of the external vector and
tensor fields $V$ and $T$. Taking $m$ to be small and switching
off the external fields, we can show that $\tilde B$ turns out to
be the same as $B$ to order $\mathcal{O}(m)$.

Averaging $\mathrm{Det}_{\mathrm{low}}$ over the instanton
collective coordinates $\xi$, which provides the partition
function $Z_N[V,T,m]$, we can obtain the fermionized representation of
the partition function in Eq.(\ref{tildeB}): 
\begin{eqnarray}
Z_N [V,T,m] &=& \langle {\mathrm{Det}}_{\mathrm{low}} [V,T,m]\rangle =
\langle \mathrm{Det} \tilde B\rangle \cr &=& \int D\psi
D\psi^{\dagger} \exp\left(\int d^4 x \psi^{\dagger}(
\rlap{/}{p} +\rlap{/}{V}+\sigma_{\mu\nu} T_{\mu\nu}+ im
)\psi\right) \prod^{N_{\pm}}
W_{\pm}[\psi^{\dagger} ,\psi ],  \label{part-func}
\end{eqnarray}
where
\begin{eqnarray}
W_{\pm}[\psi^{\dagger} ,\psi]&=& \int d^4\xi_{\pm}\prod_{N_f}\int
d^4 x \left(\psi^{\dagger} (x)\,\bar L^{-1}(x,\xi_\pm)\,
\rlap{/}{p} \Phi_{\pm , 0} (x; \xi_{\pm})\right)\cr &\times& \int
d^4 y\left(\Phi_{\pm , 0}^\dagger (y; \xi_{\pm} ) (\rlap{/}{p}\,
L^{-1}(y,\xi_\pm) \psi (y)\right). \label{tildeV}
\end{eqnarray}
The fermion fields $\psi^{\dagger} ,\psi$ are interpreted as the dynamical
quark fields or constituent quark fields induced by the zero-modes of the
instantons.

Note that the partition function of Eq.(\ref{part-func}) is
invariant under local flavor rotations due to the gauge connection
$L$ in the interaction term
 $W_{\pm}[\psi^{\dagger} ,\psi ]$.  While we preserve the gauge
invariance of the effective chiral action by introducing the gauge
connection, we have to pay a price: The effective action depends
on the path in the gauge connection $L$. We will choose the
straight-line path, though there is in general no physical reason
why other choices should be excluded. However, in
Refs.~\cite{Kim:2004hd,Goeke:2007bj} it was shown explicitly that
for the magnetic susceptibility the path dependence does not come
into play.

\subsection{Chiral effective action in the presence of
external vector and tensor fields}
We are now in a position to
derive the relevant partition function for the magnetic
susceptibility of the QCD vacuum. As exposed in
Refs.~\cite{Kim:2004hd,Goeke:2007bj} we first have to average the
low-frequency part of the quark determinant
$\mathrm{Det}_{\mathrm{low}}$ over collective coordinates
$\xi_\pm$. Here we assume a distribution of instanton sizes with
vanishing width, i.e. of the form $d(\rho)=\delta(\rho-\bar\rho)$.
Having integrated over collective coordinates and having made a
exponentiation, we reduce the partition function to the following
form (for the case $N_f=2$):
\begin{eqnarray}
Z_N&=&\int d \lambda_+d \lambda_-D\bar\psi D\psi
\exp(-\Gamma), \label{Z:withTensorTerms} \\
\Gamma &=& N_\pm \ln\frac{K}{\lambda_\pm}-N_\pm+\psi^\dagger
\left(i\rlap{/}{\partial}+\rlap{/}{V}+\sigma_{\mu\nu}T_{\mu\nu} +
im\right)\psi + \lambda _\pm Y_2^\pm, \\
Y_2^\pm&=&\alpha^2 \det_f J^\pm+\beta^2 \det_f
J^\pm_{\mu\nu}\label{Y2:definition},\\
\frac{\beta^2}{\alpha^2}&:=& \frac{1}{8N_c}\frac{2 N_c}{2
  N_c-1}=\frac{1}{8N_c-4}={\cal
  O}\left(\frac{1}{N_c}\right)\label{Tensor},\\
J^\pm&=&\psi^{\dagger '} \frac{1\pm \gamma_5}{2}\psi',\cr
J^\pm_{\mu\nu}&=&\psi^{\dagger '} \sigma_{\mu\nu}\frac{1\pm
  \gamma_5}{2}\psi'=\frac12
\left(J_{\mu\nu}\pm\frac{i}{2}\epsilon_{\mu\nu\rho\lambda}
  J_{\rho\lambda}\right),\cr
\sigma_{\mu\nu} &=&
\frac{i}{2}\epsilon_{\mu\nu\rho\lambda} \sigma_{\rho\lambda}\label{gf},\\
J_{\mu\nu}&=&\psi^{\dagger '} \sigma_{\mu\nu}\psi'\label{J},\\
\psi^{\dagger'}&=&\psi^\dagger \bar L,\;\;\psi'=L^{-1}\psi,
\end{eqnarray}
where the determinant runs over the flavor space tacitly,
and $K$ is some inessential constant making the argument
of the logarithm dimensionless.  From Eq.(\ref{Tensor})
we immediately see that the contribution of the tensor terms is just
one of the $1/N_c$-corrections.  For simplicity, we will consider the
tensor terms $J_{\mu\nu}^\pm$ later.

Having taken $N_+=N_-=N/2$, having integrated over fermion
fields~\cite{Diakonov:1985eg,Diakonov:1995qy,Musakhanov:2002xa,
Musakhanov:1998wp, Musakhanov:2001pc}, and having made a bosonization,
we end up with the partition function in the presence of the external
vector and tensor fields:
\begin{equation}
Z_N[V,T,m] =\int d\lambda D\Phi \exp\left(-\Gamma
  [V,T,m,\lambda,\Phi] \right),   \label{Z}
\end{equation}
where
\begin{equation}
\Gamma[V,T,m, \lambda,\Phi] = N\ln\frac{K}{\lambda} -N
+ \Gamma_{\Phi} + \Gamma_{\psi}  \label{S}
\end{equation}
with
\begin{eqnarray}
\Gamma_{\Phi} &=& 2\int d^4 x \Phi^2 ,   \\
\Gamma_{\psi} &=& -\mathrm{Tr} \ln \left[ \frac{\rlap{/}{P}
+\sigma_{\mu\nu} T_{\mu\nu} + im +
    i\frac{(2\pi\rho)^2\sqrt{\lambda}}{2g} \bar{L}F\Phi\cdot
    {\Gamma_\gamma}FL}{\rlap{/}{P}
    +\sigma_{\mu\nu} T_{\mu\nu} + i m }\right].
\end{eqnarray}
Here, $\Phi$ denotes the meson fields
$\Phi=\{\Phi_0,\bm\Phi\}=\{\sigma,\eta,\bm\sigma,\bm\pi\}$, and
$\Phi^2=\Phi_0^2+\bm\Phi^2=\sigma^2+\eta^2+\bm\sigma^2+\bm\pi^2$.
 The $P_\mu=p_\mu +eV_\mu$, $p_\mu =i\partial_\mu$ and $\mathrm{Tr}$
 stands for the functional trace, i.e.
$\int d^4 x \mathrm{tr}_c\mathrm{tr}_D\ tr_f$.  The $g$ is the
color factor defined as $g^2=2N_c(N_c^2-1)/(2N_c-1)$. The $F(p)$
stands for the quark form factor generated by the fermionic zero
modes. Though the explicit form of $F(p)$ is expressed in terms of
the Bessel functions \cite{Diakonov:1985eg}, we use in the present
work the dipole type of the form factor for simplicity:
\begin{equation}
  \label{eq:ff1}
  F(p) = \frac{\Lambda^2}{p^2+\Lambda^2},
\end{equation}
where $\Lambda$ is the cut-off mass with $\Lambda \sim 1/{\bar
\rho}$. In fact, the results with Eq.(\ref{eq:ff1}) are very
similar to those with the original form factor~\cite{Kim:2004hd}.

In order to calculate the correlation functions, we follow the general
effective action approaches~\cite{Coleman:1973jx,Jackiw74}.  
The effective partition function is now written as 
\begin{equation}
Z_N[V,T,m] =  \exp(-\Gamma_{\mathrm{eff}}[V,T,m,\lambda,\langle\Phi\rangle_0])
\end{equation}
with the effective action 
$\Gamma_{\mathrm{eff}}[V,T,m,\lambda,\langle\Phi\rangle_0]$,  
where vacuum average of the fields $\langle\Phi\rangle_0$ must be
taken as a solution of the vacuum:
\begin{equation}
\frac{\partial }{\partial \langle\Phi\rangle}
\Gamma_{\mathrm{eff}}[V,T,m,\lambda,\langle\Phi\rangle_0] = 0    
\end{equation}
and of the coupling $\lambda$ from the saddle-point condition: 
\begin{equation}
\frac{d}{d \lambda}  \Gamma_{\mathrm{eff}}
[V,T,m,\lambda,\langle\Phi\rangle_0] =
\frac{\partial }{\partial \lambda} \Gamma_{\mathrm{eff}}
[V,T,m,\lambda,\langle \Phi \rangle_0] = 0.      
\end{equation}
\subsection{Vacuum in the presence of external vector and tensor
  fields} 
The instanton vacuum naturally realizes S$\chi$SB, which is measured
by the quark condensate $\langle i\psi _{f}^{\dagger }\psi _{f}\rangle
_{0}$.  The strong quark-quark interaction generated by the instanton 
vacuum~(\ref{tildeV}) brings about the non-zero vacuum expectation
value of the scalar-isoscalar meson field, i.e.
$\sigma=\langle\Phi_0\rangle_0$, which can be obtained by
minimizing the effective action:
\begin{equation}
\frac{\partial \Gamma_{\mathrm{eff}}[m,\lambda,\sigma]}{\partial
\sigma}  =0 ,\,\,\,\, 
\frac{\partial \Gamma_{\mathrm{eff}}[m,\lambda,\sigma]}{\partial 
\lambda} =0  .
\label{eq:2}
\end{equation}
Expanding the effective action in one-loop meson approximation, we
obtain the following expression: 
\begin{equation}
\Gamma_{\mathrm{eff}} [m,\lambda,\sigma] = \Gamma_{0} [m,\lambda,\sigma] 
+\Gamma_{\mathrm{eff}}^{\mathrm{mes}}[m,\lambda,\sigma],
\label{eq:43}
\end{equation}
where $\Gamma_{\mathrm{eff}}^{\mathrm{mes}}[m,\lambda,\sigma]$
represents one-loop meson contribution and is expressed as follows: 
\begin{equation}
\Gamma_{\mathrm{eff}}^{\mathrm{mes}}[m,\lambda,\sigma]=\frac{1}{2}
\mathrm{Tr}  \ln\left(\left. \frac{\delta^2 \Gamma_{0}
    [m,\lambda,\Phi]}{\delta\Phi_i\delta 
    \Phi_j}\right|_{\sigma} 
  \right).
\label{onemesonloop}  
\end{equation}
Here the $\Gamma_0$ is just identical to the effective action in
Eq.(\ref{S}).  In Ref.~\cite{Goeke:2007bj}, it is extensively studied
how to solve Eqs.(\ref{eq:2}) and (\ref{onemesonloop}).  

We now find the vacuum average of the meson fields and coupling 
$\lambda$ in the presence of external fields $V$ and $T$.  Though, in
fact, their influence appears from the second order, it is enough to
consider them to the first order in order to calculate the correlation
functions.  Thus, we can safely use $\sigma$ and $\lambda$ derived
from Ref.~\cite{Goeke:2007bj}.  Then, we can straightforwardly 
derive the meson-loop effective action in the presence of the vector
and tensor fields similar to Eq. (\ref{onemesonloop}):
\begin{equation}
\Gamma_{\mathrm{eff}}[V,T, m,\lambda,\sigma]=\Gamma_{0}
[V,T,m,\lambda,\sigma] +
\Gamma_{\mathrm{eff}}^{\mathrm{mes}}[V,T,m,\lambda,\sigma],
\label{onemesonloopVT}  
\end{equation}
where
\begin{eqnarray}
\label{eq:3}
\Gamma_{\mathrm{eff}}^{\mathrm{mes}}[V,T,m,\lambda,\sigma] &=&
\frac12 \mathrm{Sp}\ln\left[4\delta_{ij}
-\frac1{\sigma^2}\mathrm{Tr}\frac{1}{\rlap{/}{P} + \sigma_{\mu\nu}
T_{\mu\nu}+i\mu(P)} \right. \cr &&\hspace{2.5cm}\times
\left.\Gamma_{i} M(P)\frac{1}{\rlap{/}{P} +\sigma_{\mu\nu} 
T_{\mu\nu}+i\mu(P)} \Gamma_{j}M(P)\right]
\end{eqnarray}
with $\mu(P) = m + M(P)$ and $M(P)=MF^2(P)$.  The $M$ denotes the
dynamical quark mass defined as
\begin{equation}
M=\frac{(2\pi\rho)^2\sqrt{\lambda}}{2g}\sigma .
\label{classical}
\end{equation}
The $\mathrm{Sp}$ stands for the trace over the meson degrees of
freedom. 

We have to keep for the self-consistency of the one-meson-loop
approximation the meson propagators in the leading
order (without meson loops) from the equation: 
\begin{equation}
\frac{\partial \Gamma_{0} [m,\lambda,\sigma]}{\partial
\sigma}  =0 .
\end{equation}
Then, the meson propagator can be redefined as
 \begin{equation}
  \label{eq:6}
{\Pi}_i(q)\Rightarrow\tilde{\Pi}_i(q) = \frac{\sigma^2
\mathbf{V}_0}{\mathrm{Tr}\mathcal{Q}(p) +
  \mathrm{Tr}\left(\mathcal{Q}(p)\Gamma_{i}\mathcal{Q}(p+q)
\Gamma_{i}\right) }.
\end{equation} 
\section{Meson-loop corrections to the magnetic 
susceptibility of the QCD vacuum}
\begin{figure}[ht]
\centering
\includegraphics[scale=0.5]{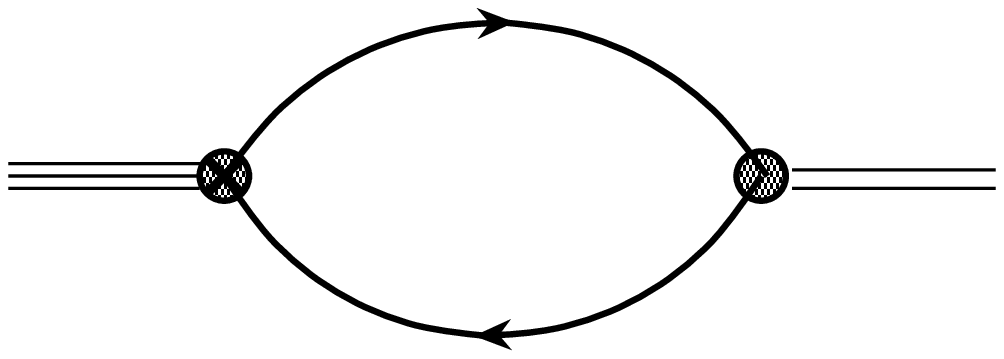} %
\includegraphics[scale=0.5]{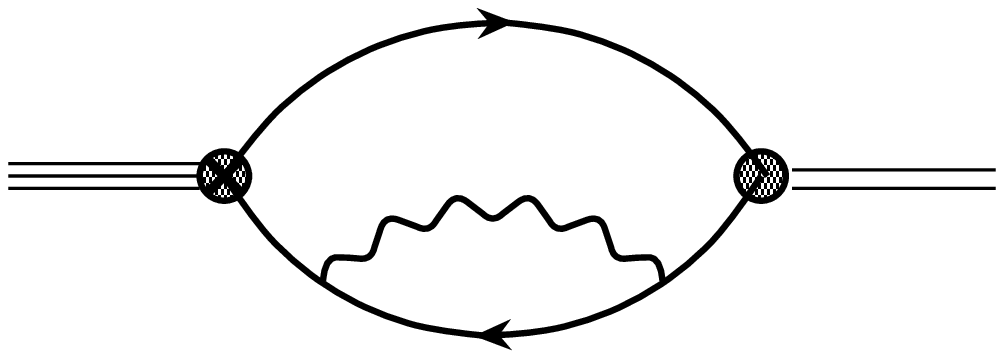} %
\includegraphics[scale=0.5]{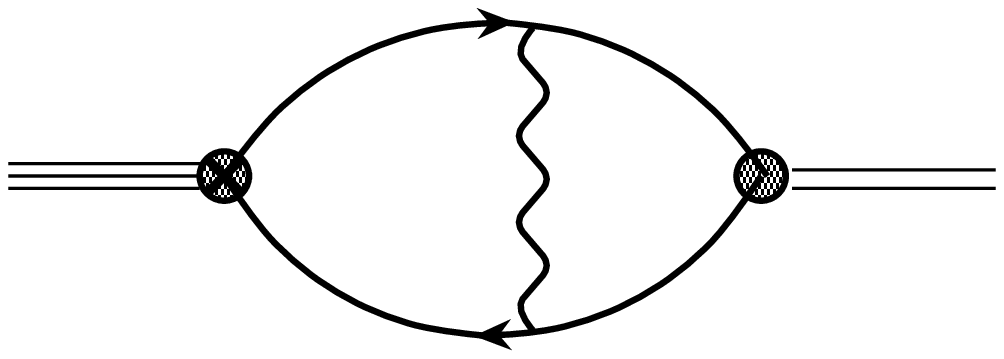}
\caption{The diagrams which contribute to the magnetic
susceptibility in the leading order (first) and next-to-leading
order (NLO) in $1/N_c$. The triple
line on the left corresponds to the local tensor current $\bar
\psi\protect \sigma_{\protect\mu\protect\nu}\psi$, the double line on
the right represents the vector current $\bar \psi\protect%
\gamma_{\protect\mu}\psi$. The interaction of the vector current
with quarks
contains both the local and nonlocal terms (see Appendix~\protect\ref%
{sec:app1} and \ref{sec:app2}).}
\label{fig:1}
\end{figure}
We are now ready for the calculation of the magnetic
susceptibility $\chi$ from the instanton vacuum, taking into
account the meson-loop corrections.  Diagrammatically, we consider
three different contributions as shown in Fig.~\ref{fig:1}.  The
first diagram in Fig.~\ref{fig:1}, which is the leading-order (LO)
contribution in the large $N_c$ expansion, has been already
calculated in Ref.~\cite{Kim:2004hd}:
\begin{equation}
\chi \langle i\psi^\dagger \psi \rangle_0 = 4 N_c \int \frac{d^4p}{%
(2\pi)^4} \left[\frac{\mu(p)}{(p^2+\mu(p)^2)^2} -
\frac{m}{(p^2+m^2)^2} \right] - 4 N_c \int \frac{d^4p}{(2\pi)^4}
\frac{pM(p)F'(p)}{(p^2+\mu(p)^2)^2},  \label{chi-final-integ}
\end{equation}
where the quark condensate in the chiral limit plays the role of
the normalization.  The infrared region of the first integral in
Eq.(\ref{chi-final-integ}) brings out the contribution of order
$\mathcal{O}(m\,\ln\,m)$ which is almost model-independent. In
order to calculate the magnetic susceptibility explicitly, we
first consider two different regions while integrating:  $0<p<1
\,\mathrm{GeV}$ and $1\,\mathrm{GeV}< p < \infty$.  Then we are
able to evaluate the part of the free quark analytically. The
second integral in Eq.~(\ref{chi-final-integ}) is related to the
nonlocal contribution without which the vector current is not
conserved, i.e. the Ward-Takahashi identity is broken.  It arises
from the nonlocal quark-quark interaction and is called
\emph{nonlocal current}.

The meson-loop corrections can be derived by differentiating the
mesonic effective action given in Eq.(\ref{eq:3}) with respect to the
external tensor field:
\begin{eqnarray}
  \label{eq:mesonc1}
\left.\frac{\delta \Gamma_{\mathrm{eff}}^{\mathrm{mes}}}{\delta
  T_{\mu\nu}(z)}\right|_{T=0} &=&  -\left(\frac{M_0}{\sigma}\right)^2
\sum_i\int d^4 x
d^4y\,\Pi_i(x-y)\mathrm{Tr}\left[\left(F(P)\frac{1}{\rlap{/}{P} +
      i\mu(P)}\right)_{yz}   \sigma_{\mu\nu} \right. \cr
&&\times\left.
  \left(\frac{1}{\rlap{/}{P} + i\mu(P)}F(P)\right)_{zx}   \Gamma_i
\left(F(P)\frac{1}{\rlap{/}{P}+i\mu(P)}F(P)\right)_{xy}
  \Gamma_i\right],
\end{eqnarray}
which corresponds to the second and third diagrams in
Fig.~\ref{fig:1}.  Note that here the order of the operators
should be kept properly, since $[P_\mu, P_\nu]=ie F_{\mu\nu}$.
Using the Schwinger method~\cite{Schwinger:nm,Vainshtein:xd}, we
continue to calculate Eq.(\ref{eq:mesonc1}).  Having carried out a
laborious but straightforward calculation, we end up with the
expression for the meson-loop corrections to the magnetic
susceptibilities:
\begin{equation}
  \label{eq:mesonc2}
\chi \langle i\psi^\dagger \psi\rangle_0^{\mathrm{mes}}=\frac{1}{2}\sum_i
\int \frac{d^4q}{(2\pi)^4}
V_i^{\chi}(q) \tilde{\Pi}_i(q),
\end{equation}
where $V_i^{\chi}$ are the vertex functions which are given
explicitly in Appendix A and B. The $\Pi_i(q)$ denote the meson
propagators given in Eq.(\ref{eq:6}).  The final expression for
the magnetic susceptibility is now written as
\begin{eqnarray}  \label{chi_total}
&&\chi\langle \bar\psi\psi\rangle=4 N_c\int \frac{d^4p}{(2\pi)^4}\frac{%
\mu(p)- p M F(p)F^{\prime }(p)}{(p^2+\mu^2(p))^2}-  \notag \\
&&4 N_c\int
\frac{d^4p}{(2\pi)^4}\frac{m}{(p^2+m^2)^2}+\frac{1}{2}\sum_i \int \frac{%
d^4q}{(2\pi)^4} V_i^{\chi}(q) \tilde \Pi_i(q).
\end{eqnarray}

The numerical result for Eq.(\ref{chi_total}) is finally obtained as
follows:
\begin{eqnarray}  \label{eq:chi_num}
\chi\langle\bar\psi\psi\rangle_{0} &=& N_c
\left[0.015\,\mathrm{GeV} + 5.3\times
  10^{-4}\,\left(\frac{m}{\mathrm{GeV}}\right) + \frac{m}{
2\pi^2}\,\ln\left(\frac{m}{\mathrm{GeV}}\right) \right] \cr &-&
\left[0.007\,\mathrm{GeV} - \left(\frac{0.415
m}{\mathrm{GeV}}\right) - \,\left(\frac{0.198
m}{\mathrm{GeV}}\right)\, \ln\left(\frac{m}{\mathrm{GeV}}
\right)\right]\cr &+& \mathcal{O}\left(m^2,\frac{1}{N_c^2}\right)
\cr &=& 0.038\,\mathrm{GeV} - \,\left(\frac{0.413
    m}{\mathrm{GeV}}\right)
-\left(\frac{0.0462 m}{\mathrm{GeV}}\right)\,\ln\left(
  \frac{m}{\mathrm{GeV}} \right) \cr
&+& \mathcal{O}\left(m^2,\frac{1}{N_c^2}\right).
\end{eqnarray}
\section{Contribution of the tensor terms to the magnetic susceptibility of the
  QCD vacuum}
Since the meson-loop corrections, however, are of order
$\mathcal{O}(1/N_c)$ in the large $N_c$ expansion, we have to take
into account the tensor terms in Eq.(\ref{Y2:definition}) which are
also of $1/N_c$ order.  It is rather tedious to consider the tensor
terms~\cite{Goeke:2007bj}.  In the present Section, we want to show
briefly how to associate with the tensor terms for the magnetic
susceptibility.  The details can be found in Ref.~\cite{Goeke:2007bj}.

\begin{figure}[ht]
\includegraphics[scale=0.8]{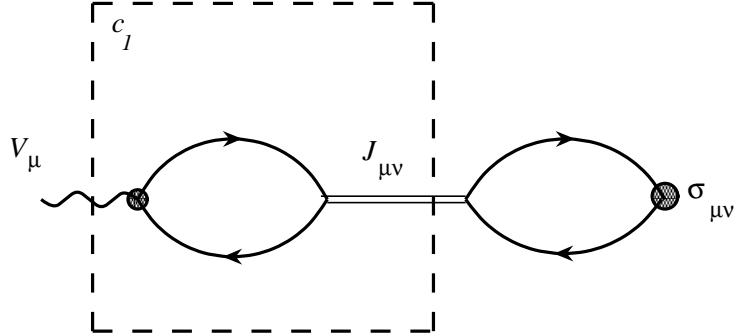}
\caption{The Feynman diagram for the
  tensor contribution.}
\label{fig:2}
\end{figure}

Introducing the tensor meson fields, the bosonized effective partition
function can be written as follows:
\begin{equation}
  \label{eq:z1}
Z_N[V,T,m]=\int d\lambda d\lambda D \Phi D \Phi_{\mu\nu} \exp
(-\Gamma[V,T,m,\lambda,\Phi,\Phi_{\mu\nu}] ),
\end{equation}
where
\begin{eqnarray}
\Gamma &=&- N_\pm
\ln\lambda_\pm+2\left(\Phi_i^2+\frac12\Phi_{\mu\nu,i}^2\right)
 \cr
&-&  \mathrm{Tr} \ln \left[\rlap{/}{P} +
  T_{\mu\nu}\sigma_{\mu\nu} + im + i \sqrt{\lambda}\bar L
  F(p)\left(\alpha \Phi_i\Gamma_i+\frac12\beta
    \Phi_{\mu\nu,i}\sigma_{\mu\nu}\Gamma_i\right) F(p)
  L^{-1}\right]\label{EffAct2}.
\end{eqnarray}
Then, the gap equation and meson fields are changed as follows:
\begin{eqnarray}
\frac{N}{\mathbf{V}_0}&=&\frac{1}{2}\mathrm{Tr}\left(\frac{i
\sqrt{\lambda} \bar
    L F(p)(\alpha \Phi_i\Gamma_i+\frac12\beta
    \Phi_{\mu\nu,i}\sigma_{\mu\nu}\Gamma_i) F(p) L^{-1}}{\rlap{/}{P}
    + im + i \sqrt{\lambda}\bar L F(p)(\alpha \Phi_i\Gamma_i+\frac12\beta
    \Phi_{\mu\nu,i}\sigma_{\mu\nu}\Gamma_i) F(p)
    L^{-1}}\right), \label{GapLambda}\\
 \Phi_i&=&\frac14\mathrm{Tr}\left(\frac{i \sqrt{\lambda}\bar L
     F(p)\alpha\Gamma_i F(p)
    L^{-1}}{\rlap{/}{P}+im + i \sqrt{\lambda}\bar L F(p)(\alpha
    \Phi_i\Gamma_i+\frac12\beta \Phi_{\mu\nu,i}\sigma_{\mu\nu}\Gamma_i)
    F(p) L^{-1}}\right), \label{GapSigma}\\
 \Phi_{\mu\nu,i}&=&\frac14\mathrm{Tr}\left(\frac{i \sqrt{\lambda}\bar L
    F(p)\beta\sigma_{\mu\nu}\Gamma_i F(p) L^{-1}}{\rlap{/}{P}+im + i
    \sqrt{\lambda}\bar L F(p)(\alpha \Phi_i\Gamma_i+\frac12\beta
    \Phi_{\lambda\rho,i}\sigma_{\lambda\rho}\Gamma_i) F(p).
    L^{-1}}\right)
\label{GapSigmaMuNu}
\end{eqnarray}
In general, it is of great difficulty to evaluate
Eqs.(\ref{GapLambda})-(\ref{GapSigmaMuNu}).  However, if $V_\mu$
varies very slowly, the calculation becomes much simpler.  Indeed,
for $V_\mu=\mathrm{const}.$ in Eq.(\ref{GapSigmaMuNu}), we get,
respectively,
\begin{equation}
\Phi_{\mu\nu}\approx c_1 iF_{\mu\nu}+\mathcal{O}(V^3)\label{c1} .
\end{equation}
The constant $c_1$ is proportional to the coupling of the tensor meson
to the vector current $V_\mu$, which is just the inverse of the
tensor-meson propagator, as will be shown below.  Then, we can show
that saddle-point values of $\lambda$ and $\sigma$ acquire only ${\cal
  O}(V^2)$-order corrections, but not ${\cal O}(V)$.  Substituting
these results into the effective action Eq.(\ref{EffAct2}), one
can easily get
\begin{equation}
\left.\frac{\partial \Gamma}{\partial
  T_{\mu\nu}}\right|_{T=0} = \mathrm{Tr}\left(\sigma_{\mu\nu}\frac{1}{\rlap{/}{P}
+i\,m + i M\bar L F(p)(1+\frac{\beta}{2\alpha \sigma}
    \Phi_{\lambda\rho,i}\sigma_{\lambda\rho}\Gamma_i) F(p)
    L^{-1}}\right)
\label{eq:tens2}
\end{equation}
with
\begin{equation}
\sigma=\sqrt{\frac{N}{2\mathbf{V}_0}}+{\cal O}(V^2)  .
\end{equation}
It is then straightforward but tedious to calculate the
contribution of the tensor terms to the magnetic susceptibility:
\begin{equation}
\chi \langle i\psi^\dagger\psi\rangle_0 = 4 N_c \int \frac{d^4p}{(2\pi)^4}
\frac{\mu(p)-p M F(p) F'(p)}{(p^2+\mu^2(p))^2}+
4 N_c \frac{\beta c_1}{\alpha \sigma} \int \frac{d^4p}{(2\pi)^4} \frac{M
  F^2(p)\mu^2(p)}{(p^2+\mu^2(p))^2}     \label{chiFinalt}
\end{equation}
with
\begin{equation}
c_1=\frac{2 N_c
  \frac{\beta}{\alpha\sigma}}{\left(1- 2N_c\left(\frac{\beta}{\alpha
        \sigma}\right)^2 \int \frac{d^4p}{(2\pi)^4}\frac{M^2
      F^4(p)\mu^2(p)}{(p^2+\mu^2(p))^2}
\right)} \int \frac{d^4p}{(2\pi)^4} \frac{M
  F^2(p)(\mu(p)-p M F(p) F'(p) )} {(p^2+\mu^2(p))^2}. \label{c1Final}
\end{equation}
In Fig.~\ref{fig:2}, the contribution of the tensor terms is drawn in
the dashed box.

The numerical results for the contribution of the tensor term are as
follows:
\begin{equation}
\chi \langle i\psi^\dagger \psi \rangle_0^{\mathrm{Tensor}} =
0.28\times 10^{-3}\,\mathrm{GeV},  \;\;\; \frac{\chi \langle
i\psi^\dagger \psi\rangle_0^{\mathrm{Tensor}}}{\chi  \langle i
\psi^\dagger \psi\rangle_{LO}}  \simeq 0.006. \label{eq:finalt}
\end{equation}
Thus, the contribution of the tensor terms to the magnetic
susceptibility turns out to be below $1\,\%$, so that we safely
neglect them.
\section{Finite-width effects on the magnetic susceptibility of the
  QCD vacuum}
We have assumed so far that the width of the instanton-size
distribution can be approximated to zero, i.e.:
\begin{equation}
d(\rho)=\delta(\rho-\bar\rho), \label{zerowidth}
\end{equation}
where all instantons have the same size $\bar\rho$.  This
approximation is justified in the large $N_c$ limit as shown in
Refs.~\cite{Diakonov:1983hh,Diakonov:1985eg}:
\begin{equation}
\frac{\langle\rho^2\rangle-\langle\rho\rangle^2}{\langle\rho\rangle^2}\sim
\mathcal{O}\left(\frac{1}{N_c}\right).
\label{SizeDefinition}
\end{equation}
Since we, however, consider the $1/N_c$ corrections to the
magnetic susceptibility of the QCD vacuum, we also have to take
into account the effects of the finite width, which are of order
$\mathcal{O}(1/N_c)$ as well. For the numerical evaluation, we
take the value for the fluctuation of the width as
\begin{equation}
\delta\rho^2=\langle\rho^2\rangle-\langle\rho\rangle^2
\approx\frac{0.56}{N_c}\,\mathrm{GeV}^{-2}, \label{SizeNumber}
\end{equation}
which arises from the two-loop size distribution of
Ref.\cite{Diakonov:1983hh}.  In order to take into account the
effects of the finite width, we first return to
Eq.~(\ref{tildeV}), in the evaluation of which
Eq.~(\ref{zerowidth}) was assumed. The integration over $\rho$
does not change the exponentiation procedure so that we still can
obtain the standard $2N_f$ quark-quark interaction in the
effective action $\Gamma$. However, we should slightly modify the
standard procedure to carry out the bosonization. For any function
$J(z,\rho)$ we have the identity
\begin{equation}
\exp\left(-\int d^4 z d\rho J^2(z,\rho)\right) = \int
D\Phi\exp\left(-\frac{1}{4}\int d^4 z d\rho\,
  \Phi^2 + \int d^4zd\rho\, \Phi J\right).
\end{equation}
Thus, the effective action is written as
\begin{eqnarray}
\Gamma &=&\frac{N}{\mathbf{V}_0} \ln \lambda + 2\int d^4 z d\rho\,
\Phi^2(z,\rho)\\
&-&\mathrm{Tr}\ln\left(\rlap{/}{P}+\sigma_{\mu\nu}T_{\mu\nu} + i m+i
  c\, \sqrt{\lambda} \int d^4 z d\rho \hat
  K(z,\rho)\Phi(z,\rho)\right), \label{EffActSize}
\end{eqnarray}
where $c$ is an inessential constant and $\hat K_{x,y}(z,\rho)\simeq
\bar \phi(x-z,\rho)\phi(z-y,\rho)$.  The LO gap equations are then
expressed as
\begin{eqnarray}
\frac{N}{\mathbf{V}_0}&=&\frac12  \mathrm{Tr} \left(\frac{i c\,
    \sqrt{\lambda}\int d^4 z d\rho \Phi(\rho)\hat
    K(z,\rho)}{(\rlap{/}{p}+i m +i c\,
    \sqrt{\lambda}\int d^4 z d\rho \hat
    K(z,\rho)\Phi(\rho)}\right),\label{eq:gaprn}\\
\Phi_0(\rho) &=& \frac14 \mathrm{Tr} \left(\frac{i c\, \sqrt{\lambda}
    \hat  K(z,\rho)}{\rlap{/}{p} + i m +i c\,
    \sqrt{\lambda} \int d^4 z d\rho \hat
    K(z,\rho)\Phi_0(\rho)}\right),\label{Gap:FWC}
\end{eqnarray}
where $\Phi_0(\rho)$ has the quantum numbers of the $\sigma$. The
unknown variables in Eqs.(\ref{eq:gaprn},\ref{Gap:FWC}) are the
parameter $\lambda$ and the function $\Phi(\rho)$. In general,
these are very complicated equations but can be solved numerically.
However, when the width is small, we can expand them with respect to
$\delta\rho^2$.  In particular, we expand the functions $\Phi(\rho)$
and $\hat K(\rho)$ with respect to $(\rho-\bar\rho)$, so that we
obtain a system of equations for the Taylor coefficients $\Phi_i$ and
$\lambda$.  Solving it, we derive the function $\Phi(\rho)$ (at
least in the vicinity of $\bar\rho$).  Details of the evaluation can
be found in Ref.~\cite{Goeke:2007bj}.   Finally, we are able to
evaluate the corrections of the finite width of the instanton size:
\begin{equation}
(\chi\langle \bar\psi \psi\rangle_0)^{\mathrm{FW}}= \left(-0.00033
+ 0.011 \left(\frac{m}{\mathrm{GeV}}\right) -
  0.096\left(\frac{m}{\mathrm{GeV}}\right)^2\right).
\label{eq:fwfinal}
\end{equation}
The numerical calculation of Eq.(\ref{eq:fwfinal}) turns out to be very
tiny.  The effects of the finite width is about $0.6\%$.  Thus, we
can safely neglect these effects on the magnetic susceptibility.
\section{Results and Discussion}
\begin{figure}[ht]
\centering
\includegraphics[scale=1.0]{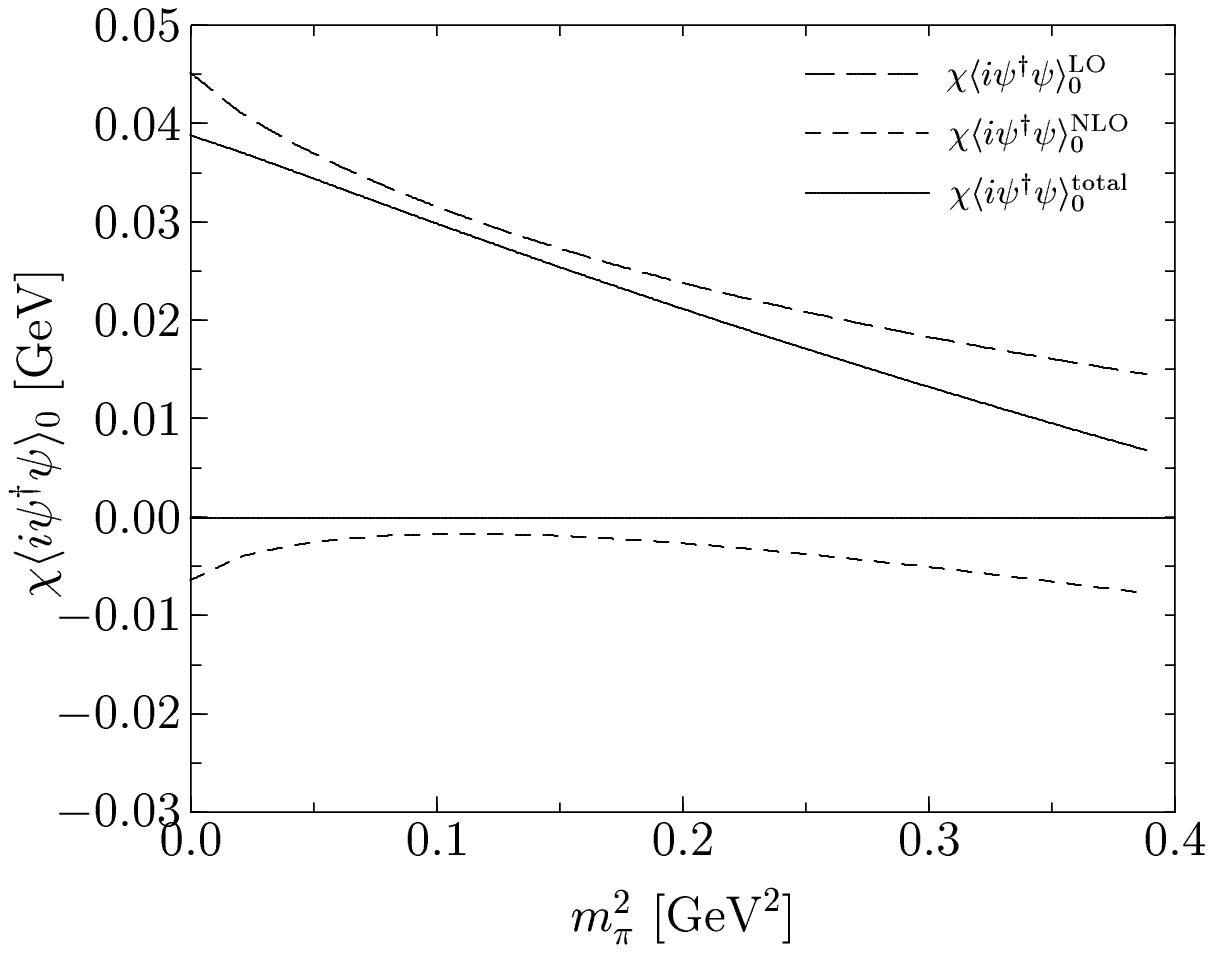}
\caption{The dependence of the magnetic susceptibility $\protect\chi%
\langle i \psi^\dagger \psi\rangle_0$ on the square of the pion
mass, $m_\pi^2$. The long-dashed curve is a contribution of the
leading order result (LO), the short-dashed curve is that
of the $1/N_c$ -correction (NLO), and the solid curve is
the total contribution.  Note that in the given range the total
result depends on $m_\pi^2$ almost linearly: The chiral-log terms
in the leading-order and next-to-leading order results almost
cancel each other.} \label{fig:re}
\end{figure}

We have calculated the magnetic susceptibility $\chi\langle
i\psi^\dagger \psi\rangle_0$ as a function of the current quark
mass, with the form factor in Eq.~(\ref{eq:ff1}) employed. For the
further discussion it is more appropriate to plot the magnetic
susceptibility in terms of the square of the pion mass, $m_\pi^2$.
Following Ref. \cite{Goeke:2007bj} both quantities are directly
related as
\begin{eqnarray}
m_\pi^2&=&m\left(\left(3.49+\frac{1.63}{N_c}\right)+
m\left(15.5+\frac{18.25}{N_c}+\frac{13.5577}{N_c} \ln m
\right)+{\cal O}(m^2)\right)\cr &=&
m\,(4.04\,+21.587\,m+4.52\,m\ln m +{\cal
O}(m^2))\,[\mathrm{GeV}^2].
\label{Res:Mpi}
\end{eqnarray}
Figure~\ref{fig:re} shows the magnetic susceptibility $\chi\langle
i\psi^\dagger \psi\rangle_0$ as a function of the square of the
pion mass, $m_\pi^2$, with the form factor in Eq.~(\ref{eq:ff1})
employed.
The long-dashed curve is the result of the leading-order
contribution in the large $N_c$ expansion, which is the same as
that in Ref.~\cite{Kim:2004hd}.  The short-dashed curve depicts
the meson-loop corrections expressed in Eq.(\ref{eq:mesonc2}).  It
is negative for all values of $m_\pi^2$.  Moreover, since the
chiral-log term in the leading-order contribution is almost
cancelled by the meson-loop corrections, the total magnetic
susceptibility depends on $m_\pi^2$ almost linearly, as shown in
Fig.~\ref{fig:re}.

The meson-loop corrections suppress the magnetic susceptibility by
about 15 $\%$ near the physical value of $m_\pi^2=(0.14\,
\mathrm{GeV})^2$$=0.02$ $\mathrm{GeV}^2$. This fact implies that
the large $N_c$ expansion seems to be reliable for them.  As a
result, the magnetic susceptibility for the up and down quarks is:
$\chi_{u,d}=35\sim 40$ MeV. As for the up and down quarks, the
present result is comparable to that of
Ref.~\cite{Belyaev:ic,Balitsky:aq,Ball:2002ps}.

Actually, the effect of the meson-loop corrections depends on
$m_\pi^2$. As one can see in Fig.~\ref{fig:re} at $m_\pi^2 \sim
0.4$ $\mathrm{GeV}^2$ they bring the leading order term down to
almost $50\%$. Thus it makes no sense to calculate the
susceptibility for larger values of $m_\pi^2$. Even this limited
range given in Fig.~\ref{fig:re} might be interesting for
practitioners of lattice gauge QCD calculations. Those
calculations suffer generally under the fact that by technical
reasons they cannot be performed with current quark masses
corresponding to the physical pion mass. Usually they have to use
$m_\pi^2>(0.4 \sim 0.6)\,\mathrm{GeV}^2$. In order to obtain a
physical observable one has to extrapolate from the $m_\pi^2$ used
in the lattice calculation to $m_\pi^2=(0.14\, \mathrm{GeV})^2$.
There exist different extrapolation techniques as e.g. linear
extrapolation, extrapolation by chiral perturbation theory
\cite{Bernard:2003rp,Procura:2006bj,Procura:2005ev,Frink:2004ic,Frink:2005ru},
finite range extrapolation
~\cite{Leinweber:2003dg,Leinweber:2005zj}, extrapolation by
suitable chiral models
\cite{Leinweber:1998ej,Goeke:2005fs,Goeke:2007fq}, etc. Apparently
the present calculations suggest in Fig.~\ref{fig:re} that for the
magnetic susceptibility of the vacuum a simple linear
extrapolation might be sufficient.

\section{Summary and Conclusion}
In the present work, we have investigated the magnetic
susceptibility of the QCD vacuum, based on the low-energy
effective QCD partition function from the instanton vacuum.  We
first have constructed the effective partition function in the
presence of the external vector and tensor fields as well as of
the current quark mass.  We have made the effective action
gauge-invariant at the outset.  We also have considered the
meson-loop corrections which are a part of the $1/N_c$
corrections. To be consistent, we have taken into account the
effects of the finite width of the instanton size distribution as
well as the tensor terms of the quark-quark interactions, since
they are also of order $\mathcal{O}(1/N_c)$.  However, both these
effects turn out to be negligibly small.

The meson-loop corrections contribute to the magnetic
susceptibility $\chi$ negatively and their chiral-log terms almost
cancel those of the leading-order contribution.  They contribute
to the magnetic susceptibility for the up and down quarks by
around $15\,\%$. As a result, assuming for the up quarks (and
identically for the down quarks) $m_{u,d}\simeq 5\,\mathrm{MeV}$
we obtain the magnetic susceptibility $\chi\langle i \psi^\dagger
\psi\rangle_0 = 35\sim 40 \,\mathrm{MeV} $.  These results are
comparable to the estimates of the vector dominance and of the QCD
sum rule approaches~\cite{Belyaev:ic,Balitsky:aq,Ball:2002ps}. It
turns out that $\chi$ is almost a linearly decreasing function of
the current quark mass $m$ or of $m_\pi^2$. To know this might be
useful for the chiral extrapolation of lattice data for the
magnetic susceptibility of the QCD vacuum.

\acknowledgments
The present work is supported by the Korea Research Foundation
Grant funded by the Korean Government(MOEHRD)
(KRF-2006-312-C00507) and by the bilateral funds
(DFG-436-USB-113/11/0-1) between Germany and Uzbekistan. The work
is also supported by the Transregio-Sonderforschungsbereich
Bonn-Bochum-Giessen, the Verbundforschung (Hadrons and Nuclei) of
the Federal Ministry for Education and Research (BMBF) of Germany,
the Graduiertenkolleg Bochum-Dortmund of the DFG, the COSY-project
J\"ulich as well as the EU Integrated Infrastructure Initiative
Hadron Physics Project under contract number RII3-CT-2004-506078.

\begin{appendix}
\section{Scalar vertices\label{sec:app1}}
We present the explicit expression for the vertex function for the
isoscalar scalar field $V_\sigma^{\chi}(k)$. The vertex function
for the isovector scalar field $V_{\bm\sigma}^{\chi}(k)$ is given
as $V_{\bm\sigma}^{\chi}(k)=-3V_\sigma^{\chi}(k)$.  The
$V_\sigma^{\chi}(k)$ is written as the following expression:
\begin{footnotesize}
\begin{eqnarray}
V_\sigma^{\chi}(k)&=& N_c\int\frac{d^4p}{(2\pi)^4} 
\frac{M^2 F(p) F(k+p)}{
3 p \left(p^2+\mu(p)^2\right)^3 
\left((k+p)^2+\mu\left(k+p\right)^2\right)^2} 
\left[12 p F(p) F(k+p) \left((k+p)^2+\mu\left(k+p\right)^2\right)\right.\cr
&&\times \left(-3 \mu(p) p^2-\mu(k+p) p^2+2 M F(p) 
\left(p^2-\mu(p) \mu(k+p)+k\cdot p\right) F'(p) p \right. \cr
&& + \left. 2 \mu(p)^2 \mu(k+p)-3 \mu(p) k\cdot p\right)+ 
\frac{2 F(p) F(k+p) \left(p^2+\mu(p)^2\right)}{(k+p)} \left(
 -3 \mu(k+p) p^4 - 3 k \mu(k+p) p^3 \right. \cr
&& - 6 \mu(k+p) k\cdot p p^2-6 k \mu(k+p) k\cdot p p+
 M \left((k+p) F(p) \left(-3 p^4+2 k^2 p^2-6 \mu(k+p)^2 p^2
   \right.\right. \cr
&& \left.\left.-9 k\cdot p p^2-8 (k\cdot p)^2\right) F'(p)-
 p F(k+p) \left(3 p^4+k^2 p^2+6 k\cdot p p^2+2
   (k\cdot p)^2\right) F'(k+p)\right) \cr
&& + \mu(p) \left(-3 p^4-3 k p^3-3 k\cdot p p^2- 6 M
   (k+p) F(p) \mu(k+p) F'(p) p^2+6 (k+p) \mu(k+p)^2 p \right. \cr
&&\left. - 3 k pk\cdot p + 4 M^2 F(p) F(k+p) \left(k^2 p^2-(k\cdot
    p)^2\right) F'(p) F'(k+p)\right)\cr
&&\left.-3 (k+p) \mu(p)^2 \left(M \left(F(p) \left(p^2+k\cdot p\right) F'(p)+p
   \sqrt{(k+p)^2} F(k+p) F'(k+p)\right)-2 p
   \mu(k+p)\right)\right)\cr
&& + 4 \left(p^2+\mu(p)^2\right)
   \left((k+p)^2+\mu\left(k+p\right)^2\right) \left(-p \left(3
       \mu(p)-2 M p F(p) F'(p)\right)\right. \cr 
&& \left.\times  \left(\frac1{p^3}\left(F(k+p) \left(p F''(p) (k\cdot p)^2+\left(k^2
          p^2+k\cdot p p^2-(k\cdot p)^2\right)
        F'(p)\right)\right)\right.\right.\cr 
&&- \left. \frac1{(k+p)^3}\left(F(p) \left((k+p) \left(k^2+k\cdot
   p\right)^2 F''(k+p) 
-\left(k^4+3 (k\cdot p)^2+\left(3 k^2+p^2\right)
   k\cdot p\right) F'(k+p)\right)\right)\right)\cr
&& -  \left(3p \mu(p)+3p\mu(k+p)+2 M F(p) k\cdot p F'(p)\right)
\left(F(k+p) \left(p 
   F'(p)+k\cdot p F''(p)\right)\right.\cr
&&- \frac1{(k+p)^3}\left(F(p) \left( (k+p) (k^2+k\cdot p)
(p^2+k\cdot p) F''(k+p)\right.\right. \cr
&& \left.\left.\left.\left.\left.- \left(p^4+2 k^2 p^2+(k\cdot
   p)^2+\left(k^2+3 p^2\right) k\cdot p\right)
   F'(k+p)\right)\right)\right)\right)\right].
\end{eqnarray}  
\end{footnotesize}

\section{Pseudoscalar vertices\label{sec:app2}}
We present the explicit expression for the vertex function of the
pseudoscalar isoscalar field $V_\eta^{\chi}(k)$. The vertex
function of the pseudoscalar isovector field
$V_{\bm\pi}^{\chi}(k)$ is given as
$V_{\bm\pi}^{\chi}(k)=-3V_\eta^{\chi}(k)$.  The
$V_\eta^{\chi}(k)$ is written as the following expression: 
\begin{footnotesize}
\begin{eqnarray}
V_\eta^{\chi}(k)&=& N_c\int\frac{d^4p}{(2\pi)^4} 
\frac{M^2 F(p) F(k+p)}{3 p \left(p^2+\mu(p)^2\right)^3
   \left((k+p)^2+\mu\left(k+p\right)^2\right)^2} \left[-12 p F(p) F(k+p)
   \left((k+p)^2+\mu\left(k+p\right)^2\right) \right.\cr
&& \left(-3 \mu(p) p^2+\mu(k+p) p^2+2 M F(p) \left(p^2+\mu(p)
    \mu(k+p)+k\cdot p\right) F'(p) p-2 \mu(p)^2 \mu(k+p)-3 \mu(p)
  k\cdot p\right) \cr
&&+\frac{2 F(p) F(k+p)    \left(p^2+\mu(p)^2\right)
}{(k+p)}\left(3 \mu(k+p) p^4+3 k \mu(k+p) p^3+6 
   \mu(k+p) k\cdot p p^2+6 k \mu(k+p) k\cdot p p\right.\cr
&& + M  \left((k+p) F(p) \left(-3 p^4+2 k^2 p^2-6 \mu(k+p)^2 p^2-9
    k\cdot p p^2-8 (k\cdot p)^2\right) F'(p)\right.\cr
&&+\left. p F(k+p) \left(3 p^4+k^2 p^2+6 k\cdot p p^2+2
   (k\cdot p)^2\right) F'(k+p)\right)-\mu(p) 
\left(3 p^4+3 k p^3+3 k\cdot p p^2\right.\cr
&& -6 M (k+p) F(p) \mu(k+p) F'(p) p^2-6 (k+p) \mu(k+p)^2 p+3 k
   k\cdot p p \cr
&&\left.+ 4 M^2 F(p) F(k+p) \left(k^2 p^2-(k\cdot p)^2\right) F'(p)
   F'(k+p)\right)\cr
&& \left.+3 (k+p) \mu(p)^2 \left(M \left(p (k+p) F(k+p)
   F'(k+p)-F(p) \left(p^2+k\cdot p\right) F'(p)\right)-2 p
   \mu(k+p)\right)\right)\cr
&&+ 4 \left(p^2+\mu(p)^2\right)
   \left((k+p)^2+\mu\left(k+p\right)^2\right)\cr
&& \left(p \left(3 \mu(p)-2 M
   p F(p) F'(p)\right) \left(\frac1{p^3}\left(F(k+p) \left(p F''(p)
       (k\cdot p)^2+\left(k^2 p^2+k\cdot p 
   p^2-(k\cdot p)^2\right) F'(p)\right)\right)\right.\right.\cr
&&-\left.\frac1{(k+p)^3}\left(F(p) \left((k+p) \left(k^2+k\cdot
   p\right)^2 F''(k+p)-\left(k^4+3 (k\cdot p)^2+\left(3 k^2+p^2\right)
   k\cdot p\right) F'(k+p)\right)\right)\right)\cr
&&+  \left(3p \mu(p)-3p  \mu(k+p)+2 M F(p) k\cdot p F'(p)\right)
\left(F(k+p) \left(p F'(p)+k\cdot p F''(p)\right)\right.\cr
&&-\frac1{(k+p)^3}\left(F(p) \left(\sqrt{(k+p)^2} (k^2+k\cdot
    p) (p^2+k\cdot p) F''(k+p) \right.\right.\cr 
&& \left.\left.\left.\left.\left. -(p^4+2 k^2 p^2+(k\cdot p)^2+ (k^2+3
          p^2) k\cdot p)  F'(k+p)\right)\right)\right)\right)\right].
\end{eqnarray}
\end{footnotesize}

\end{appendix}


\begin{thebibliography}{99}
\bibitem{Ioffe:1983ju}
  B.~L.~Ioffe and A.~V.~Smilga,
  Nucl.\ Phys.\  B {\bf 232}, 109 (1984).

\bibitem{Belyaev:ic} V.~M.~Belyaev and Y.~I.~Kogan,
Yad.\ Fiz.\ \textbf{40}, 1035 (1984). 

\bibitem{Balitsky:aq} I.~I.~Balitsky, A.~V.~Kolesnichenko and A.~V.~Yung,
Sov.\ J.\ Nucl.\ Phys.\ \textbf{41}, 178 (1985)  [Yad.\ Fiz.\
\textbf{41}, 282 (1985)]. 


\bibitem{Ball:2002ps} P.~Ball, V.~M.~Braun and N.~Kivel,
Nucl.\ Phys.\ B \textbf{649}, 263 (2003). 


\bibitem{Petrov:1998kg} V.~Y.~Petrov, M.~V.~Polyakov, R.~Ruskov, C.~Weiss
and K.~Goeke,
Phys.\ Rev.\ D \textbf{59}, 114018 (1999). 


\bibitem{Kim:2004hd}  H.-Ch.~Kim, M.~M.~Musakhanov and M.~Siddikov,
Phys.\ Lett.\ B \textbf{608}, 95 (2005).  


\bibitem{Braun:2002en} V.~M.~Braun, S.~Gottwald, D.~Y.~Ivanov, A.~Sch\"afer
and L.~Szymanowski,
Phys.\ Rev.\ Lett.\ \textbf{89}, 172001 (2002).



\bibitem{Kim:2005jc}  H.-Ch.~Kim, M.~M.~Musakhanov and M.~Siddikov,
Phys.\ Lett.\ B \textbf{633}, 701 (2006).  



\bibitem{Goeke:2007bj}  K.~Goeke, M.~M.~Musakhanov and M.~Siddikov,
arXiv:0707.1997 [hep-ph].  


\bibitem{Shuryak:1981ff} E.~V.~Shuryak,
Nucl.\ Phys.\ B \textbf{203}, 93 (1982). 


\bibitem{Diakonov:1983hh} D.~Diakonov and V.~Y.~Petrov,
Nucl.\ Phys.\ B \textbf{245}, 259 (1984). 


\bibitem{Diakonov:2002fq} D.~Diakonov, 
Prog.\ Part.\ Nucl.\ Phys.\ \textbf{51}, 173 (2003) 
[arXiv:hep-ph/0212026].



\bibitem{Schafer:1996wv} T.~Sch\"afer and E.~V.~Shuryak,
Rev.\ Mod.\ Phys.\ \textbf{70}, 323 (1998)  [arXiv:hep-ph/9610451].


\bibitem{Chu:vi} M.~C.~Chu, J.~M.~Grandy, S.~Huang and J.~W.~Negele,
Phys.\ Rev.\ D \textbf{49}, 6039 (1994)  [arXiv:hep-lat/9312071].


\bibitem{Negele:1998ev} J.~W.~Negele,
Nucl.\ Phys.\ Proc.\ Suppl.\ \textbf{73}, 92 (1999)  [arXiv:hep-lat/9810053].


\bibitem{DeGrand:2001tm} T.~DeGrand,
Phys.\ Rev.\ D \textbf{64}, 094508 (2001) [arXiv:hep-lat/0106001].


\bibitem{Faccioli:2003qz} P.~Faccioli and T.~A.~DeGrand,
Phys.\ Rev.\ Lett.\ \textbf{91}, 182001 (2003) [arXiv:hep-ph/0304219].


\bibitem{Bowman:2004xi} P.~O.~Bowman, U.~M.~Heller, D.~B.~Leinweber,
A.~G.~Williams and J.~b.~Zhang,
Nucl.\ Phys.\ Proc.\ Suppl.\ \textbf{128}, 23 (2004) [arXiv:hep-lat/0403002].


\bibitem{Cristoforetti:2006ar}  M.~Cristoforetti, P.~Faccioli, M.~C.~Traini
and J.~W.~Negele,
Phys.\ Rev.\ D \textbf{75}, 034008 (2007) [arXiv:hep-ph/0605256].



\bibitem{Musakhanov:2002xa} M.~M.~Musakhanov and H.-Ch.~Kim,
Phys.\ Lett.\ B \textbf{572}, 181 (2003) [arXiv:hep-ph/0206233].


\bibitem{Musakhanov:1996qf} M.~M.~Musakhanov and F.~C.~Khanna,
Phys.\ Lett.\ B \textbf{395}, 298 (1997) [arXiv:hep-ph/9610418].


\bibitem{Salvo:1997nf} E.~D.~Salvo and M.~M.~Musakhanov,
Eur.\ Phys.\ J.\ C \textbf{5}, 501 (1998) [arXiv:hep-ph/9706537].

\bibitem{Nam:2006sx}
  S.~i.~Nam and H.-Ch.~Kim,
  Phys.\ Rev.\  D {\bf 74}, 076005 (2006)
  [arXiv:hep-ph/0609267].

\bibitem{Ryu:2006bf}
  H.~Y.~Ryu, S.~i.~Nam and H.-Ch.~Kim,
  arXiv:hep-ph/0610348.

\bibitem{Nam:2007fx}
  S.~i.~Nam and H.-Ch.~Kim,
  Phys.\ Rev.\  D {\bf 75}, 094011 (2007)
  [arXiv:hep-ph/0703089].

\bibitem{'tHooft:1976fv}  G.~'t Hooft,  Phys.\ Rev.\ D \textbf{14},
  3432 (1976) [Erratum-ibid.\ D \textbf{18}, 2199 (1978)]. 

\bibitem{Lee:sm} C.~k.~Lee and W.~A.~Bardeen,
Nucl.\ Phys.\ B \textbf{153}, 210 (1979). 


\bibitem{Diakonov:1985eg} D.~Diakonov and V.~Y.~Petrov,
Nucl.\ Phys.\ B \textbf{272}, 457 (1986). 


\bibitem{Diakonov:1995qy} D.~Diakonov, M.~V.~Polyakov and C.~Weiss,
Nucl.\ Phys.\ B \textbf{461}, 539 (1996)  [arXiv:hep-ph/9510232].

\bibitem{Musakhanov:vu}  M.~Musakhanov,  Nucl.\ Phys.\ A \textbf{699} (2002)
340.

\bibitem{Musakhanov:1998wp} M.~Musakhanov,
Eur.\ Phys.\ J.\ C \textbf{9}, 235 (1999) [arXiv:hep-ph/9810295].

\bibitem{Musakhanov:2001pc} M.~Musakhanov,
Nucl.\ Phys.\ A \textbf{699}, 340 (2002)  
[arXiv:hep-ph/0104163]. 

\bibitem{Coleman:1973jx}
S.~R.~Coleman and E.~Weinberg, Phys.\ Rev.\ D {\bf 7}, 1888 (1973). 

\bibitem{Jackiw74}
R.~Jackiw, Phys.\ Rev.\ D {\bf 9}, 1686 (1974).


\bibitem{Schwinger:nm} J.~S.~Schwinger,
Phys.\ Rev.\ \textbf{82}, 664 (1951). 


\bibitem{Vainshtein:xd} A.~I.~Vainshtein, V.~I.~Zakharov, V.~A.~Novikov and
M.~A.~Shifman,
Sov.\ J.\ Nucl.\ Phys.\ \textbf{39}, 77 (1984)  [Yad.\ Fiz.\
\textbf{39}, 124 (1984)]. 

\bibitem{Bernard:2003rp}
  V.~Bernard, T.~R.~Hemmert and U.~G.~Meissner,
  Nucl.\ Phys.\ A {\bf 732}, 149 (2004) 
  [arXiv:hep-ph/0307115].


\bibitem{Procura:2006bj}
  M.~Procura, B.~U.~Musch, T.~Wollenweber, T.~R.~Hemmert and W.~Weise,
  Phys.\ Rev.\  D {\bf 73}, 114510 (2006) 
  [arXiv:hep-lat/0603001].

\bibitem{Procura:2005ev}
  M.~Procura, B.~U.~Musch, T.~R.~Hemmert and W.~Weise,
  Nucl.\ Phys.\ Proc.\ Suppl.\  {\bf 153}, 229 (2006) 
  [arXiv:hep-lat/0512026].

\bibitem{Frink:2004ic}

  M.~Frink, U.~G.~Meissner and I.~Scheller,
  Eur.\ Phys.\ J.\ A {\bf 24}, 395 (2005) 
  [arXiv:hep-lat/0501024].

\bibitem{Frink:2005ru}
  M.~Frink, U.~G.~Meissner and I.~Scheller,
  Eur.\ Phys.\ J.\ A {\bf 24}, 395 (2005) 
  [arXiv:hep-lat/0501024].

\bibitem{Leinweber:2003dg}
  D.~B.~Leinweber, A.~W.~Thomas and R.~D.~Young,
  Phys.\ Rev.\ Lett.\  {\bf 92}, 242002 (2004) 
  [arXiv:hep-lat/0302020].

\bibitem{Leinweber:2005zj}
  D.~B.~Leinweber, A.~W.~Thomas and R.~D.~Young,
  Lect.\ Notes Phys.\  {\bf 663}, 113 (2005).


\bibitem{Leinweber:1998ej}
  D.~B.~Leinweber, D.~H.~Lu and A.~W.~Thomas,
  Phys.\ Rev.\ D {\bf 60}, 034014 (1999) 
  [arXiv:hep-lat/9810005].


\bibitem{Goeke:2005fs}
  K.~Goeke, J.~Ossmann, P.~Schweitzer and A.~Silva,
  Eur.\ Phys.\ J.\  A {\bf 27},  77 (2006)
  [arXiv:hep-lat/0505010].

\bibitem{Goeke:2007fq}
  K.~Goeke, J.~Grabis, J.~Ossmann, P.~Schweitzer, A.~Silva and D.~Urbano,
  Phys.\ Rev.\  C {\bf 75}, 055207 (2007)
  [arXiv:hep-ph/0702031].



\end{thebibliography}
\end{document}